%% file: ms.tex
\documentclass[conference]{IEEEtran}
\usepackage{amssymb}
\usepackage{amsmath}
\usepackage{bm}
\usepackage{graphicx}
\usepackage{xcolor}
\usepackage{import}
\usepackage[utf8]{inputenc}
\usepackage{listings}
\usepackage[noend]{algpseudocode}
\usepackage{url}
\usepackage{multirow}
\usepackage[font={small,it}]{caption}
\usepackage[bookmarks=false]{hyperref} 
\usepackage{array}
\usepackage{rotating}   
\usepackage{makecell}
\usepackage{colortbl}
\usepackage{enumerate}
\usepackage{booktabs}
\usepackage{subfig}
\usepackage{paralist}
\usepackage[ruled,vlined]{algorithm2e}
\usepackage{mathptmx}
\newcommand{\subparagraph}{} 
\setdefaultleftmargin{0em}{0em}{}{}{}{}

\let\url\nolinkurl

\algrenewcommand\algorithmiccomment[1]{\hfill \textcolor{gray}{$\triangleright$ \textit{#1}}}

\DeclareUnicodeCharacter{00A0}{ }

\makeatletter
\g@addto@macro\normalsize{%
  \setlength\abovedisplayskip{4pt}
  \setlength\belowdisplayskip{4pt}
  \setlength\abovedisplayshortskip{4pt}
  \setlength\belowdisplayshortskip{4pt}
}
\makeatother

\renewcommand{\tablename}{Tab.}

\renewcommand{\figurename}{Fig.}
\renewcommand{\paragraph}[1]{\vspace{0.1cm}\noindent{\bf #1.}}

\newcommand{\mysubsubsection}[1]{\subsubsection{#1}}

\ifCLASSINFOpdf
\else
\fi
\IEEEoverridecommandlockouts

\begin{document}
\title{A Security Reference Architecture for Blockchains}
\author{
	\IEEEauthorblockN{
		Ivan Homoliak\IEEEauthorrefmark{1} \\
		ihomoliak@sutd.edu.sg
	}
	\and
	\IEEEauthorblockN{
		Sarad Venugopalan\IEEEauthorrefmark{1}\\
		sarad\_venugopalan@sutd.edu.sg
	}
	\and
	\IEEEauthorblockN{
		Qingze Hum\IEEEauthorrefmark{1}\\
		qingze\_hum@mymail.sutd.edu.sg
	}
	\and
	\IEEEauthorblockN{
		Pawel Szalachowski\IEEEauthorrefmark{1}\\
		pawel@sutd.edu.sg
	}
	\and
	\IEEEauthorblockA{
		\hspace{5cm}
		\IEEEauthorrefmark{1}Singapore University of Technology and Design
	}
\thanks{$\copyright$ 2019 IEEE.  Personal use of this material is permitted.  Permission from IEEE must be obtained for all other uses, in any current or future media, including reprinting/republishing this material for advertising or promotional purposes, creating new collective works, for resale or redistribution to servers or lists, or reuse of any copyrighted component of this work in other works.}

}

\maketitle   

\begin{abstract}
	\input{sec/intro}

\end{abstract}

\section{Blockchains at a Glance}
\label{sec:background}
\input{sec/background}

\section{Security Reference Architecture}
\label{sec:model}

\input{sec/model}

\section{Network Layer}
\label{sec:network}
\input{sec/network}

\section{Consensus Layer}
\label{sec:consensus}
\input{sec/consensus}

\section{Replicated State Machine Layer}
\label{sec:smart_contracts}
\input{sec/smart_contracts}

\section{Application Layer}
\label{sec:apps}
\input{sec/apps}

\section{Conclusion}
\label{sec:conclusion}
\input{sec/conclusion}

\section*{Acknowledgment}
\input{sec/ack}

\bibliographystyle{abbrv}

\bibliography{ref}


\end{document}

%% file: sec/intro.tex
Due to their interesting features, blockchains have become popular in recent years.
They are full-stack systems where security is a critical factor for their
success.
The main focus of this work is to  systematize knowledge about security and privacy issues of blockchains.
To this end, we propose a security reference architecture based on models that
demonstrate the stacked hierarchy of various threats (similar to the
ISO/OSI hierarchy) as well as threat-risk assessment
using ISO/IEC 15408.
In contrast to the previous
surveys~\cite{conti2018survey,bano2017consensus,wang2018survey,2015-Bitcoin-SOK},
we focus on the categorization of security incidents based on their origins and using the proposed architecture we present existing prevention and mitigation techniques.
The scope of our work mainly covers aspects related to decentralized nature of blockchains, while we mention common operational security issues and countermeasures only tangentially.

%% file: sec/background.tex
The blockchain is a data structure representing an append-only distributed ledger
that consists of entries (a.k.a., transactions) aggregated within ordered blocks.
The order of the blocks is agreed by untrusting participants running a consensus
protocol.
A transaction is an elementary data entry that may contain arbitrary data, e.g., an order to transfer native cryptocurrency (i.e., crypto-tokens), a piece of application code (i.e., smart contract), the execution orders of such application code, etc. 
Transactions sent to a blockchain are validated by all nodes that maintain a
replicated state of the blockchain.

\begin{figure}[t]
      \centering        
      \includegraphics[width=0.94\columnwidth]{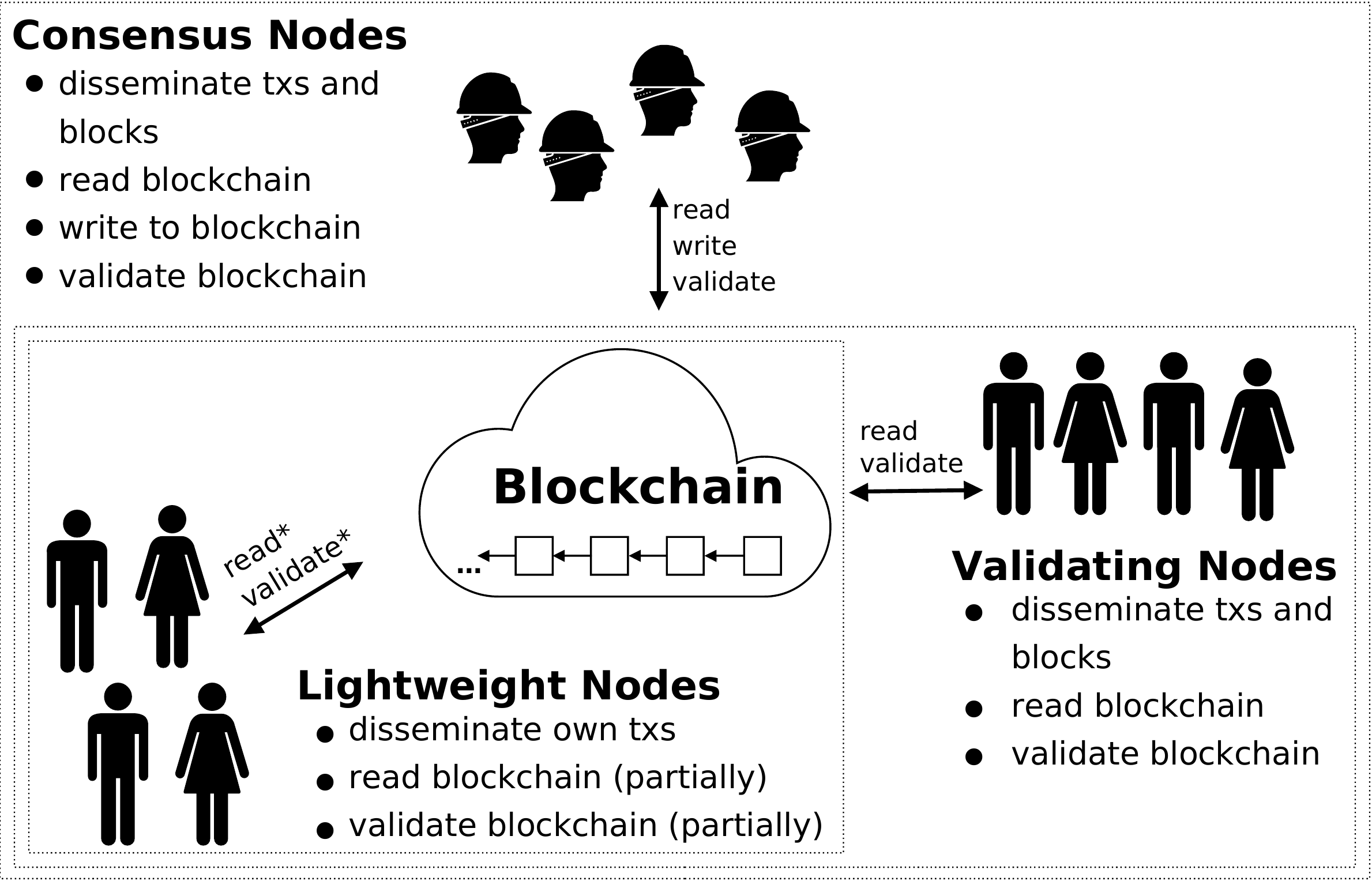}         
      \caption{Involved parties with their interactions and hierarchy.}
      \label{fig:node-types}
  \end{figure}

\paragraph{Involved Parties}
Blockchains usually involve the following parties (see~\autoref{fig:node-types}).

    \textit{(1)~Consensus nodes} actively participate in the underlying consensus protocol. 
    These nodes can read the blockchain and write to it by appending new transactions.
    Besides, they can validate the blockchain and thus check whether writes of other consensus nodes are correct and respect a specified logic. 
    By a combination of writing and validation capabilities,
    consensus nodes can prevent malicious behavior (e.g., by not appending invalid transactions, or not following an incorrect blockchain view).
    These nodes disseminate transactions to be appended within a block to the blockchain.
    In the context of Proof-of-Resource protocols (see \autoref{sec:PoR}), these nodes are often referred to as \textit{miners}. 
    \textit{(2)~Validating nodes} read the entire blockchain, validate it, and disseminate transactions to be appended to the blockchain. 
    Unlike consensus nodes, validating nodes cannot write to the blockchain. 
    Thus, they cannot prevent malicious behavior. 
    However, since they possess copies of the entire blockchain, they can detect malicious behavior.
    \textit{(3)~Lightweight nodes} (a.k.a., clients) benefit from most of the blockchain functionalities, but they are equipped only with limited information about the blockchain.  
    These nodes read only a fragment of the blockchain (usually block headers) and validate only a small number of transactions that concern them, while they rely on consensus and validating nodes for ensuring correctness of the blockchain. 
    Therefore, they can detect only a limited set of attacks, usually pertaining to their own transactions. \
    
\paragraph{Features of Blockchains}
Blockchains were initially proposed as open cryptocurrencies, but due to their features, they became appealing for other applications as well. 
     Blockchains achieve \textit{decentralization} via a distributed consensus protocol, which provides resilience to failures.   
     Usually, participants are equal and no single entity pose an authority.
     Another important result of decentralization is censorship resistance.
    The ledger is \textit{immutable}, requiring a significant quorum of colluding nodes to change its entries retrospectively.  Usually, immutability is achieved thanks to a cryptographic one-way function that creates integrity preserving links between blocks.
    Although blockchains are highly redundant in a storage of the data,
    the main advantage of such redundancy is high
    \textit{availability}.  
    This feature is of special interest to applications that cannot tolerate outages. 
    Blockchain transactions, as well as actions of protocol participants, are usually \textit{transparent} to other participants and in most cases even to the public. 
    This can be a benefit for multiple applications, but it can also
    be seen as a disadvantage from the anonymity and privacy perspective.
    
    Beside the features that are common in blockchains, some blockchains may focus on additional features, such as energy efficiency~\cite{gilad2017algorand,bentov2016cryptocurrencies,kiayias2017ouroboros}, scalability~\cite{nakamoto2008bitcoin}, throughput~\cite{buchman2018tendermint,Kokoris-KogiasJ18-omniledger,ZamaniM018-rapidchain}, privacy~\cite{sasson2014zerocash}, accountability~\cite{hyperledger1}, etc.
    
\paragraph{Types of Blockchains}
Based on how a new node enters a consensus protocol, we distinguish the following blockchain types.
    \textit{(1)~Permissionless} blockchains allow anyone to join the consensus
    protocol without permission.
    Such participation can be anonymous, and these protocols are
    designed to run over the Internet.
    To prevent Sybil attacks~\cite{douceur2002sybil}, these schemes usually require consensus nodes to establish their identities by running a Proof-of-Resource scheme, while the consensus power of a node is proportional to its resources invested into running the protocol.
    \textit{(2)~Permissioned} blockchains require a consensus node to obtain permission (and identity) to join the consensus protocol.
    The permission is given by a centralized or federated authority(ies), while nodes usually have equal consensus power (i.e., one vote per node).
    These schemes can be \textit{public} if they are accessible over the Internet or \textit{private} when they are deployed over a restricted network.
    \textit{(3)~Semi-Permissionless} blockchains require each new-coming consensus node to obtain a permission (i.e., cryptocurrency ``stake''); however, such permission can be given by any stakeholder (i.e., consensus node).
    These blockchains are similar
    to permissionless blockchains, except a consensus is based on a
    stake rather than on resources spent.  
    The node's consensus power is proportional to the stake it has. 
    Similar to permissionless blockchains, these systems are usually intended to be run over the Internet.
%
    Novel and interesting aspects of (semi-)permissionless blockchains are
    incentives and network effects that are designed to increase the protocol's security,
    deployability, and adoption.

%% file: sec/model.tex
\begin{figure}[t]
	\vspace{-0.3cm}
    \centering
    \includegraphics[width=0.43\textwidth]{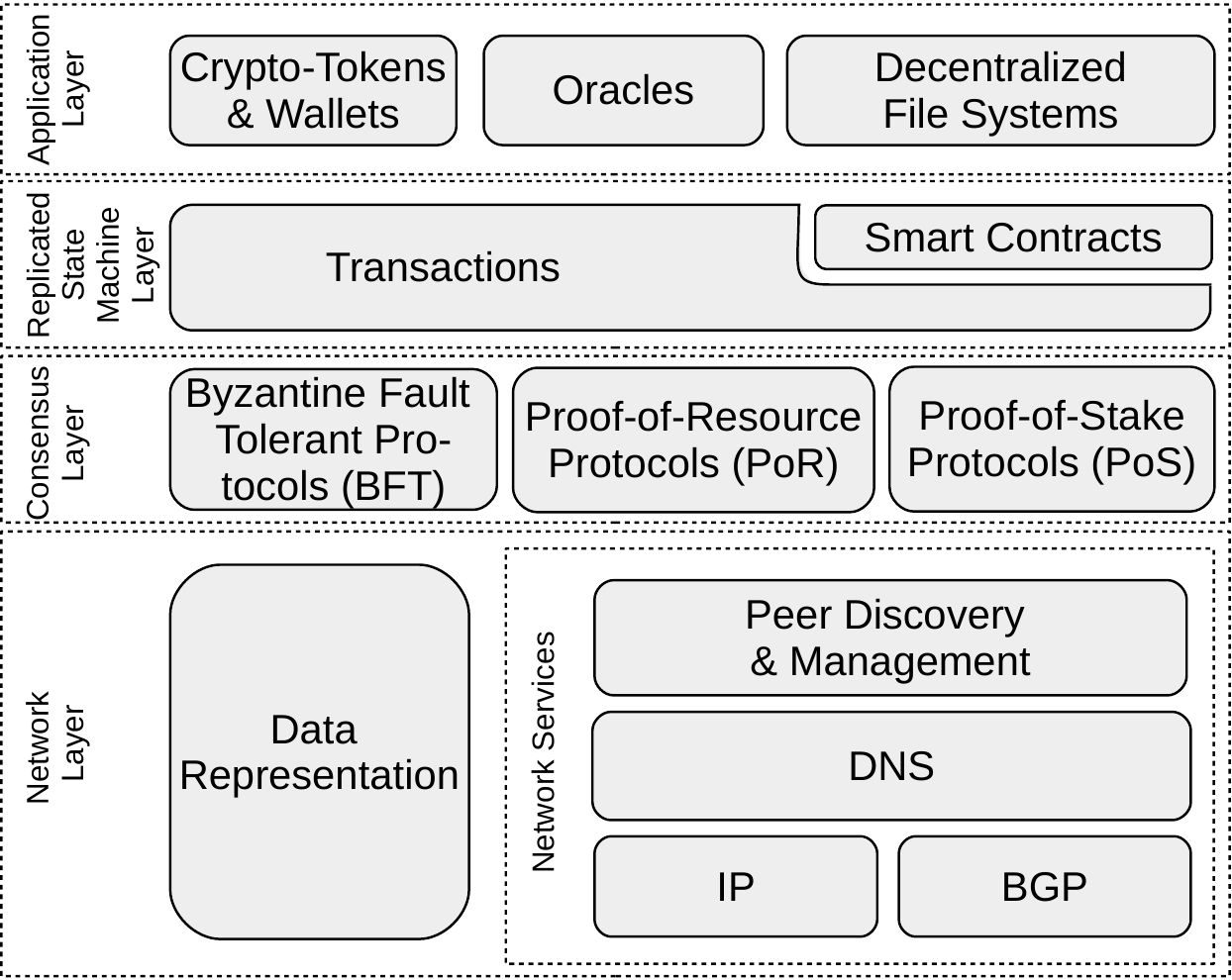}         
    \caption{Stacked model of reference architecture.}
    \label{fig:overview}
\end{figure}

\paragraph{Stacked Model}
\label{sec:LayeredModel}
To classify security aspects related to blockchains, we
introduce a simplified stacked model~\cite{wang2018survey} consisting of four layers (see \autoref{fig:overview}).  
In contrast to previous work~\cite{wang2018survey}, we preserve only such granularity level that enables us to isolate
various nature of security threats.

(1) The \textit{network layer} (see \autoref{sec:network}) consists of the data
representation and network services planes. 
The data representation plane deals with storing and encoding of data, while the
network service plane contains discovery and communication with protocol peers, addressing, routing, and naming services. 
(2) The \textit{consensus layer} (see \autoref{sec:consensus}) deals with ordering of transactions and we divide it according to a type of the protocol used to Byzantine Fault Tolerant (e.g.,~\cite{castro1999practical,bessani2014state,cachin2009ByzantinePaxos,duan2014bchain,miller2016honey}), Proof-of-Resource (e.g.,~\cite{nakamoto2008bitcoin,Dziembowski_proofsof,slimcoin,miller2014permacoin,filecoin}), and Proof-of-Stake (e.g.,~\cite{bentov2016cryptocurrencies,kiayias2017ouroboros}) protocols. 
(3) The \textit{replicated state machine (RSM) layer} (see \autoref{sec:smart_contracts}) deals with the interpretation of transactions, according to which, the state of the blockchain is updated.
Smart contracts involve two special types of transactions, which represent a programming code itself and invocations of this code together with input data.
(4) In the application layer (see \autoref{sec:apps}) we present the most common end-user functionalities such as crypto-tokens with wallets that store secrets, oracles that represent data feeds, and decentralized file systems.
Throughout the paper, we summarize components of particular layers with their respective security threats and protection techniques.

\begin{figure}[t]
	\vspace{-0.3cm}
	\centering
	\includegraphics[width=0.42\textwidth]{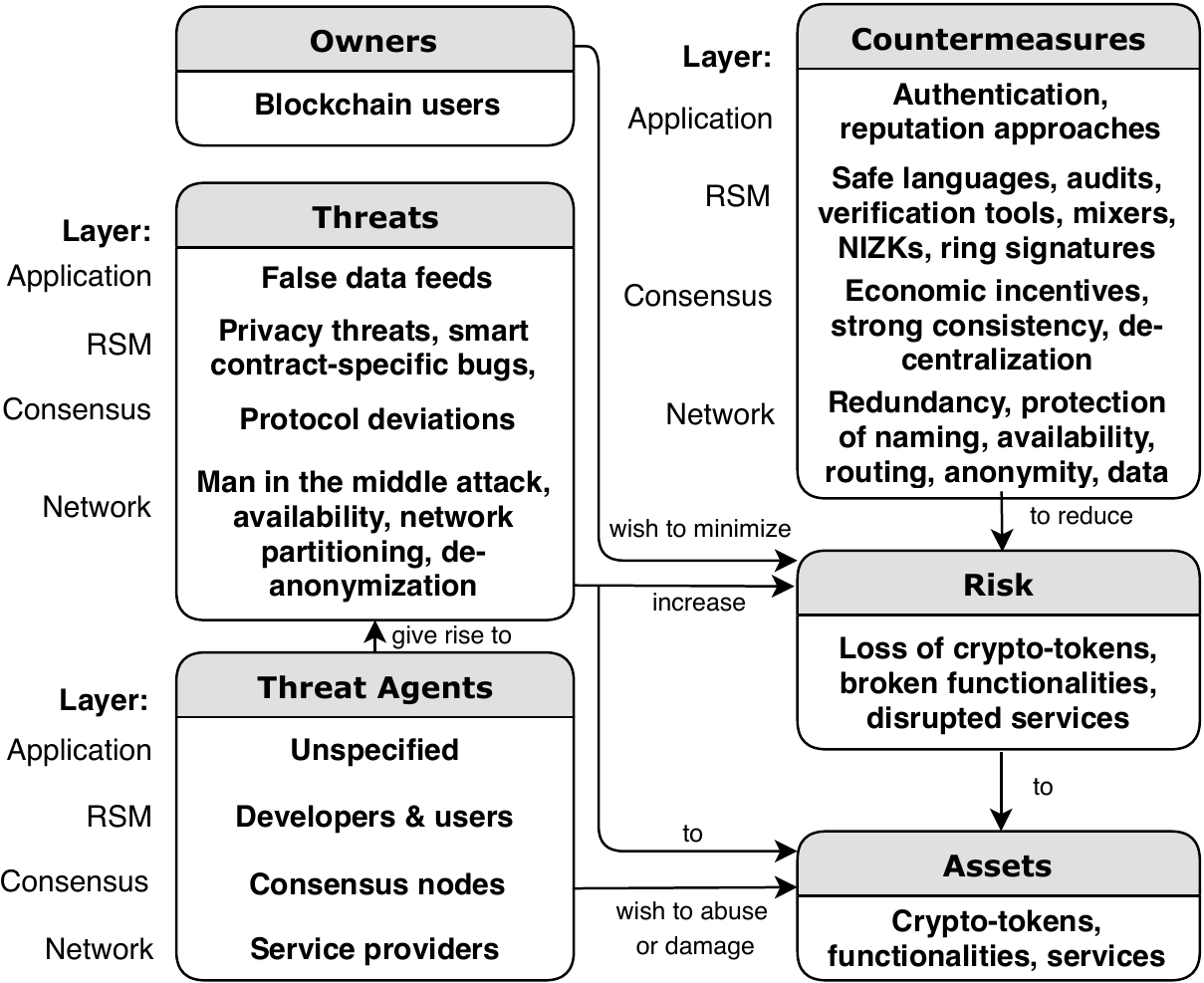}
	\caption{Threat-risk assessment model of reference architecture.}
	\label{fig:iso15408}
\end{figure}     

\paragraph{Threat-Risk Assessment Model}
To better capture security-related aspects of blockchain systems, we
introduce a threat-risk model
based on the template of ISO/IEC 15408~\cite{cc2017}. The model includes the
following components and actors (see \autoref{fig:iso15408}).
    \textit{Owners} are blockchain users who run any node type.
    Owners posses crypto-tokens and/or use blockchain-based applications or services.
    \textit{Assets} consist of monetary value (i.e., crypto-tokens),
    blockchain functionalities, as well as services built on top of them (e.g., exchanges, secure logging, supply chains). 
    \textit{Threat agents} are malicious users whose intention is to steal assets, break functionalities, or disrupt services.
    \textit{Threats} arise from vulnerabilities at the network, in smart
    contracts, from consensus protocol deviations, violations of protocol assumptions, or application-specific dependencies.
    Threats facilitate various attacks on assets and services.
    \textit{Countermeasures} are provided by the security, safety, incentives, and reputation techniques that protect owners from threats.
    \textit{Risks} caused by threats and their agents may lead to losses of monetary assets or service malfunctions and disruptions.
    
The owners wish to minimize the risk caused by threats that arise
from threat agents.
With the stacked model, different threat agents appear at each layer.  
 At the network layer, there are service providers including parties
 managing IP addresses and DNS names.
 The threats at this layer come from man-in-the-middle (MITM) attacks, 
 network partitioning, de-anonymization, and availability attacks. 
 Countermeasures contain protection of availability, naming, routing,
 anonymity, and data.
 At the consensus layer, nodes may be malicious and wish to alter the outcome of the consensus protocol by deviating from it.
 The countermeasures include economic incentives, strong consistency, and decentralization.
 At the RSM layer, the threat agents may stand for developers who (un)intentionally introduce semantic bugs in smart contracts  (intentional bugs represent backdoors).\footnote{Note that semantic bugs may occur at each layer; however, we focus only on smart contract-specific bugs that we put into the RSM layer.} 
 Mitigating countermeasures are safe languages, static/dynamic verification, and audits.
 Other threats are related to privacy of data and identity of users with mitigation techniques using mixers, privacy-preserving cryptography constructs (e.g., non-interactive zero-knowledge proofs (NIZKs), ring signatures).
 At the application layer, threat agents are unspecified, since any user on the network who uses a blockchain application may pose a threat.
 The threats on this layer arise from false data feeds and examples of mitigation techniques are authentication or reputation systems.

%% file: sec/network.tex
Blockchains are \textit{overlay networks} on top of other networks; hence, blockchains inherit security and privacy issues from the underlying networks. 
Based on permission to join the blockchain system, the networks are either private or public. 
A private network is a  network of local devices whose access is insulated from public networks. 
The Internet is a public network of interconnected autonomous systems (ASes) that relay network traffic at their borders.
The network layer is divided into data representation and network sub-planes (see \autoref{fig:overview}). 
Data representation plane is protected by cryptographic primitives that ensure data integrity, user authentication, and optionally confidentiality, privacy, anonymity, non-repudiation, and accountability. 
The main blockchain-oriented services provided by the network layer are peer management and discovery, which rely on the internals of underlying network, such as domain name resolution (i.e., DNS), network routing protocols (e.g., LAN routing for IP, WAN routing such as BGP). 
In the following, we discuss pros and cons of private and public networks and their security threats that affect overlayed blockchains.

\subsection{\textbf{Private Networks}}
\label{Private Network}
A private network ensures low latency, a centralized administration, privacy, and meeting regulatory obligations (e.g., HIPAA\footnote{Health insurance portability and accountability act \url{https://hipaa.com/}.} for healthcare data).
The organization owning the network provides access to local participants as well as to external ones when required; hence systems deploying private networks belong to the group of permissioned private blockchains. 
The inherent feature of private networks is that authentication and access control can be provided at the network layer. 

\smallskip
\subsubsection{Pros}
    
    \textit{Access control}
    is achieved by centralized authentication of users and assigning them roles. 
    A private network has full control over routing paths and physical resources used, which enables regulation of the network topology and transmission medium best suited for requirements.
    \textit{Data privacy} is ensured by permissioned settings.
    \textit{User identity}
     is revealed only within a private group of nodes. 
     They are immune to external attacks in contrast to public networks (see \autoref{ssec:InternetSecurityThreats}).    
    \textit{Fine-grained authorization controls}
    are applied by the operator of network resources to implement the security principle of minimal exposure and thus mitigate insider threat attacks on a local network. 
    \textit{Resource availability}
    is easier to manage and foresee, as all network participants and the deployment scenario are known ahead of time. 

\smallskip            
\subsubsection{Cons}

\begin{inparaenum}

    \item[\textit{Virtual Private Network (VPN)}]
    connectivity is required to communicate between private networks spread over different geographical locations. 
    While VPNs are in general secure, they inherit the disadvantages of running service over the Internet.
    \item[\textit{Applicability}]
        of private networks is suitable only for permissioned and private blockchains.
\end{inparaenum}

\mysubsubsection{Security Threats and Countermeasures}

\begin{inparaenum}

    \item[\textit{Insiders}] 
    may pose a serious threat to security~\cite{insidercorp}.
    A compromised node may already have administrative privileges or obtain them by exploiting a system, network, or security vulnerabilities. 
    \textit{Countermeasures} include regular software updates, user monitoring (e.g., SIEM~\cite{swift06}), prevention techniques that minimize trust and maximize trustworthiness, as well as respecting best practices~\cite{homoliak2019insight}.
\end{inparaenum}

\subsection{\textbf{Public Networks / the Internet}}
Public networks provide high decentralization, openness, and low entry barrier, while network latency, privacy, and network control are put aside.
These networks are naturally required by all public (permissionless) blockchain systems.

\mysubsubsection{Pros}
\begin{inparaenum}

    \item[\textit{High availability}] is attractive to multi-homed nodes since they have alternate routes to send and receive messages. 
    Multi-homed nodes may find useful to disseminate blocks across multiple channels, thereby increasing the chance of blocks being appended to the blockchain.
    \item[\textit{High decentralization}] is achieved through geographical dispersion of nodes. 
    Public peer-to-peer (p2p) networks are harder to shut down~\cite{rodrigues2010}.    
    \item[\textit{Openness and low entry barrier}] on the Internet are achieved through wide adoption, technology interoperability (e.g., using TCP/IP), economic  (e.g., low cost of broadband connection) and societal (e.g., resistance to regulations) factors~\cite{openness2016}. 
    Statistical resource sharing~\cite{mcknight1995} and openness are fundamental to a low entry barrier.
\end{inparaenum}

\mysubsubsection{Cons}\label{ssec:InternetCons}

\begin{inparaenum}

\item[\textit{Single-point-of-failure --}] 
DNS with its hierarchy, IP addresses, and ASes are managed by  centralized parties -- Internet Corporation for Assigned Names and Numbers (ICANN); in particular, Internet Assigned Numbers Authority (IANA). 
\item[\textit{External adversaries}]
pose a threat to public networks. 
These adversaries can be  classified based on their capabilities to which the  blockchain network may be exposed~\cite{internetadversary}: 
(1) resources under attacker control (e.g., botnets, DNS and BGP servers),
(2) identities are stolen or masqueraded (e.g., IP addresses participating in an eclipse attack or route manipulation),
(3) MITM attacker (i.e., eavesdropping and spoofing),
(4) common vulnerabilities leading to exploits (e.g., observed in DNS BIND software~\cite{bind2019}),
(5) revealing secrets (e.g., de-anonymizing peers).
\item[\textit{Efficiency}] 
-- although an average Internet bandwidth has improved in recent years~\cite{akamai2017}, a distribution of powerful infrastructure is not uniform, which results in a different latency among peers, and the overall latency of the network is increased -- this, in turn, may result to loss of created blocks and thus wasting of consensus power.

\end{inparaenum}
    
\mysubsubsection{Security Threats and Countermeasures}
\label{ssec:InternetSecurityThreats}

\begin{compactdesc}

\item[\textbf{DNS attacks}] commonly arise from cache poisoning~\cite{Son2010DNS} that mainly affects nodes employing DNS bootstrapping~\cite{bitcoinj} to retrieve online peers but also users of online blockchain explorers.\footnote{Note that blockchain explorers might be also affected by compromised certification authorities. Protection relying on DNS is DNS-based Authentication of Named Entities (DANE), while Certificate Transparency is mitigation relaying on centralized public logs.}
One \textit{countermeasure} is a security extension of DNS, called DNSSEC, which provides authentication and data integrity.
In addition to standard DNS, name resolution can also be made using alternate DNS servers~\cite{crispin2001}.

\item[\textbf{Routing attacks}] are traffic route diversions, hijacking, or DoS attacks.
Beside simple data eavesdropping or modification, these attacks may lead to network partitioning, which in turn raises the risks of 51\% attacks or selfish mining attacks (see \autoref{sec:consensus}). 
\textit{Countermeasures} suggest nodes to be multi-homed (or using VPN) for route diversity, choosing extra peers whose connections do not pass through the same ASes, preference of peers hosted on the same AS within the same /24 prefix (to reduce risk of partitions), and fetching the same block from multiple peers~\cite{apostolaki2017hijacking}. 
Another mitigation is SABRE~\cite{apostolaki2018sabre}, a secure relay network that runs alongside with the Bitcoin network. 
The BGPsec~\cite{rfc8205} is a security extension for BGP used between neighboring ASes, and it provides assurance of route origin and propagation by cryptographic verification.

\item[\textbf{Eclipse attacks}] hijack all connections of a node to the blockchain network.
Hence, all traffic received by the node is under the full control of the attacker.
Eclipse attacks arise from threats on DNS and routing in the network as well as they may be a result of vulnerabilities in p2p protocols~\cite{heilman2015eclipse,wust2016ethereum,marcus2018low}).
Eclipse attack increase chances of selfish mining and double spending attacks (see \autoref{sec:consensus}) -- the eclipsed victims may vote for an attacker's chain. 
\textit{Countermeasures:}
Improving randomness in choosing peers was proposed in work~\cite{heilman2015eclipse} by several rules that manage the peer table.
Another mitigation strategy against eclipse attacks is to use redundant network links or out-of-band connections to verify transactions (e.g., by a blockchain explorer).
Also, note that countermeasures for DNS and routing attacks are applicable here as well.

    \item[\textbf{DoS attacks on connectivity}] of consensus nodes may result in a loss of consensus power, thus preventing consensus nodes from being rewarded~\cite{minerddos}.
    For validating nodes, this attack leads to disruption of some blockchain dependent services~\cite{fullnodeddos}. 
    \textit{Countermeasures:}
    One mitigation is to peer only with white-listed nodes.
    Methods to prevent volumetric DDoS include on-premise filtering (i.e., with an extra network device), cloud filtering (i.e., redirection of traffic through a cloud when DDoS is detected or through a cloud DDoS mitigation service), or hybrid filtering~\cite{hybriddos} (i.e., combinations of the previous two). 

    \item[\textbf{DoS attacks on local resources,}] such as memory and storage, may reduce the peering and consensus capabilities~\cite{hdddos} of nodes.
    An example attack is flooding the network with low fee transactions (a.k.a., penny-flooding), which may cause memory pool depletion, resulting in a system crash. 
    Possible mitigation is raising the minimum transaction fee and rate-limit to the number of transactions.
    Several mitigating techniques are applied to Bitcoin~\cite{weaknessesdos} nodes including scoring DoS attacks and banning misbehaving peers.
    
    \item[\textbf{Identity revealing attacks}] are conducted by linking the IP address of a node with identity propagated in transactions~\cite{BiryukovP14,millertopology2015}. 
    Traffic analysis using Sybil listeners can reveal the linkage of node IP addresses and their transactions~\cite{chainalysis2015}.
    \textit{Countermeasures} include using VPNs or anonymization services, such as Tor.
    See \autoref{sec:privacy-threats} for further identity and privacy-protecting mechanisms at the RSM level.

\end{compactdesc}

%% file: sec/consensus.tex
The consensus layer of the stacked model deals with the ordering of transactions.
It includes three main categories of consensus protocols with regard to different principles of operation and thus their security aspects.
First, we focus on the security aspects that are generic to all categories of consensus protocols, and then we detail each category. 

\subsection{Generic Attacks}
\subsubsection{Violations of Protocol Assumptions}\label{sec:violation-of-assumptions}

\begin{compactitem}
    \item[\textbf{Adversarial Centralization of Consensus Power.}] 
    In these attacks, a design assumption about the decentralized distribution of consensus power is violated.
    Examples of this category are \textit{51\% attacks} for PoR and PoS protocols as well as
    $\frac{1}{3}$ of \textit{Byzantine nodes} for BFT protocols (and their combinations).
    %
    In a 51\% attack, the majority of the consensus power is held by the adversary, thus also the result of the protocol is under its control.
    In \textit{Byzantine attacks}, a quorum of $\frac{1}{3}$ adversarial consensus nodes might cause the protocol being disrupted or even halted.
    As a design-oriented countermeasure, it is important to promote decentralization by incentive schemes that reward honest participation and discourage~\cite{miller2015nonoutsourceable} or punish~\cite{buterin2017casper,daian2017snow} protocol violations. 
    
    \item[\textbf{Time-Validation Attacks.}]
    Usually, besides system time, nodes in PoW and PoS maintain network time that is computed as the median value of the time obtained from the peers.
    Such a time is often put into the block header, while nodes, upon receiving a block, validate whether it fits freshness constraints.
    An attacker can exploit this approach by connecting a significant number of nodes and propagate inaccurate timestamps, which can slow down or speed up the victim node's network time~\cite{timejacking}.  
    When such a desynchronized node creates a block, this block can be discarded by a network due to freshness constraints. 
    To avoid de-synchronization attacks, a node can build a reputation list of trusted peers or employ a timestamping authority~\cite{szalachowski2018short}.

\end{compactitem}

\mysubsubsection{Double Spending}
This attack is possible due to the creation of two or more conflicting blocks with the same height, resulting in inconsistencies called \textit{forks}. 
Thus, some crypto-tokens might be temporarily spent in both conflicting blocks, while only a single block is later included in the honest chain. 
To prevent this attack in permissionless blockchains, it is recommended to wait a certain amount of time (i.e., a few next blocks) until a block ``is settled.'' 

\subsection{Proof-of-Resource Protocols (PoR)}\label{sec:PoR}
Protocols from this category require nodes to prove a spending of a scarce resource in a lottery-based fashion~\cite{hyperledger1}. 
Scarce resources may stand for:
    \textit{(1) Computation}
    that is represented by Proof-of-Work (PoW) protocols (e.g., Bitcoin, Ethereum).
    \textit{(2) Storage} used in the setting of Proof-of-Space protocols~\cite{Dziembowski_proofsof} (e.g., Spacecoin~\cite{spacecoin}, SpaceMint~\cite{honsi2017spacemint}).
    \textit{(3) Crypto-tokens} spent for 
     Proof-of-Burn protocols (e.g., Slimcoin~\cite{slimcoin}).
    \textit{(4) Combinations and modification} of the previous types, such as storage and computation, called Proof-of-Retrievability (e.g., Permacoin~\cite{miller2014permacoin}) and  
    storage over time, which is represented by Proof-of-Space protocols (e.g., Filecoin~\cite{filecoin}).
    
PoR protocols belong to the first generation of consensus protocols, and they are mostly based on Nakamoto Consensus~\cite{nakamoto2008bitcoin} that utilize PoW, inheriting its pros (e.g., high scalability) and cons (e.g., low throughput). 
For the detailed analysis of several PoW designs, we refer the reader to~\cite{2019-pow-evaluation}.

\mysubsubsection{Pros}
In PoR protocols, malicious overriding of the history of blockchain (or its part) requires spending at least the same amount of resources as was spent for its creation.
This is in contrast to principles of PoS protocols, where a big enough coalition  may override the history with almost no cost.

\mysubsubsection{Cons} stand mainly for a high operational cost.
Moreover, these protocols provide only probabilistic finality, which enables attacks forking the last few blocks of the chain.  

\subsubsection{Security Threats and Mitigations}
\begin{compactitem}
    \item [\textbf{Selfish Mining:}] 
    In selfish mining~\cite{eyal2018majority}, an adversary attempts to privately build a secret chain and reveal it to the public only when an honest chain is ``catching up'' with the secret one. 
    The longest chain rule causes honest miners to adopt the attacker's chain and invalidate the honest chain, thus wasting their consensus power.
    This attack is more efficient when consensus power of a selfish miner reaches some threshold (e.g., 30\%).    
     The selfish mining strategy was later generalized~\cite{sapirshtein2016optimal} and extended to other variants that increase the profit of the attacker~\cite{nayak2016stubborn}.
    \textit{Countermeasures:} 
    (1) For the case of the longest chain rule, the first introduced mitigation is uniform tie breaking~\cite{eyal2018majority}, which tells consensus nodes to choose the chain to extend uniformly at random, regardless of which one they received first. 
    However, this technique is less effective when assuming network delays~\cite{sapirshtein2016optimal}.
    (2) As the longest chain rule enables this attack, it is recommended to use other fork choice rules that also account for the quality of solutions and make the decision deterministic, as opposed to a uniform tie breaking.
    An example of such a rule is to select the block based on the smallest hash value.
    Another example is to include partial solutions~\cite{zamyatinflux, pass2017fruitchains}  
    or full (orphaned) blocks~\cite{sompolinsky2013accelerating,zhang2017publish} for computation of block's quality.
    (3) Another option for a deterministic fork choice rule is using a pseudo-random function~\cite{kogias2016byzcoin}, which moreover provides unpredictability, hence an attacker cannot determine his chances to win a tie. 
    (4) PoW protocols can be combined with BFT protocols, where PoW is used only for joining the protocol and BFT for consensus itself (e.g., \cite{Kokoris-KogiasJ18-omniledger,ZamaniM018-rapidchain,kogias2016byzcoin}).    
    
    \item [\textbf{Feather Forking:}]
    In this attack~\cite{feather-forks}, the adversary creates incentives for rational miners to collectively censor certain transactions.
    Before a mining round begins, an adversary announces that he will not extend the block containing blacklisted transactions, and thus will attempt to extend a forked chain.
    Although this strategy is not profitable for the adversary and the success rate is dependent on his consensus power, rational honest nodes prefer to join on the censorship to avoid the potential loss.
    \textit{Countermeasures:} design-oriented protection is to minimize the chance of the attacker being successful, which can be done by including (and rewarding) partial solutions~\cite{zamyatinflux, rizun2016subchains,pass2017fruitchains} 
    or full orphaned blocks~\cite{sompolinsky2013accelerating,zhang2017publish} into branch difficulty computation.
    
    \item [\textbf{Pool Specific Attacks:}]
    Since PoR protocols are usually based on a lottery having a single winner~\cite{nakamoto2008bitcoin}, rewarding for participation imposes a high payout variance for solo miners (i.e., once in a few years).
    As a consequence, mining pools emerged and caused centralization of the mining power, which may result in selfish mining, double spending, or 51\% attacks.
    \textit{Countermeasures:}
    Non-outsourceable scratch-off puzzles~\cite{miller2015nonoutsourceable} avoid creation of pools but require each consensus node to meet high demands on connectivity and storage, as opposed to centralized pools, where only a pool operator needs to meet these demands.
    If pools are acceptable, their size can be controlled by protocols that reward partial solutions~\cite{zamyatinflux, rizun2016subchains, pass2017fruitchains} 
    and thus minimize payout variance.
    For a detailed analysis of rewarding schemes in pools, we refer the reader to~\cite{rosenfeld2011analysis}.
    
\end{compactitem}

\subsection{Byzantine Fault Tolerant (BFT) Voting Protocols}
BFT protocols represent voting-based~\cite{hyperledger1} consensus protocols that utilize Byzantine agreement and a state machine replication~\cite{schneider1990implementing}.
These protocols assume a fully connected topology, broadcasting messages, and a master-replicas hierarchy.
Synchronous examples of this category are PBFT~\cite{castro1999practical}, RBFT~\cite{aublin2013rbft}, eventually synchronous examples are BFT-SMaRt~\cite{bessani2014state}, Tendermint~\cite{buchman2018tendermint}, Byzantine Paxos~\cite{cachin2009ByzantinePaxos}, BChain~\cite{duan2014bchain},
and asynchronous examples are SINTRA~\cite{cachin2002sintra} and HoneyBadgerBFT~\cite{miller2016honey}.
For more details, we refer the reader to review of BFT protocols and their practical applications in both permissioned and permissionless blockchains~\cite{cachin2017blockchain}.

\mysubsubsection{Pros}
BFT protocols provide high throughput and a low latency finality. 
To face their scalability limitation, BFT protocols are often combined with PoS or PoW.
This is in line with a lottery approach~\cite{hyperledger1} for selecting a portion of all nodes, referred to as committee, which further runs BFT consensus (e.g., Algorand~\cite{gilad2017algorand}, Zilliqa~\cite{zilliqa2017zilliqa}, DFINITY~\cite{hanke2018dfinity}).

\mysubsubsection{Cons}
The main con of traditional BFT protocols~\cite{cachin2009ByzantinePaxos,castro1999practical} is a low scalability caused by a high communication complexity (i.e., $\Theta(n^2)$). 
Since these protocols can work efficiently only with a limited number of consensus nodes, they can be used in their pure form only in permissioned blockchains.

\mysubsubsection{Security Threats and Mitigations}

Many BFT protocols assume synchronous delivery of messages.
However, this assumption can be violated by unpredictable network scheduler, as demonstrated on PBFT protocol~\cite{miller2016honey}.
This fact motivates asynchronous BFT protocols that can be based on threshold-based cryptography, which enables reliable and consistent broadcast~\cite{cachin2002sintra,miller2016honey}.
Issues with scalability and throughput can be dealt with by applying cryptographic constructs~\cite{boneh2001short,cachin2005random,shoup2000practical} and partitioning consensus nodes into shards that process transactions in parallel~\cite{Kokoris-KogiasJ18-omniledger,ZamaniM018-rapidchain}.
Another option is to prune the number of nodes running BFT into committees~\cite{gilad2017algorand}, which, however, reduces security level of BFT and provides only probabilistic security guarantees depending on the committee size.

\subsection{Proof-of-Stake Protocols (PoS)}
Similar to the PoR category, PoS protocols are based on the lottery approach~\cite{hyperledger1}.
However, in contrast to PoR, no scarce resource is spent; instead, the nodes are required ``to prove investment'' of crypto-tokens in order to participate in a protocol, and thus potentially earn interest from the invested amount.
The concept of PoS was first time proposed in Peercoin~\cite{peercoin} as a combination with PoW -- each node has its particular difficulty for PoW, which is based on the age of the coins a node owns. 
Although there exist a few pure PoS protocols (e.g., Chains of Activity~\cite{bentov2016cryptocurrencies}, Ouroboros~\cite{kiayias2017ouroboros}), the trend is to combine them in a hybrid setting with PoR (e.g., Proof-of-Activity~\cite{bentov2014proof}, Peercoin~\cite{peercoin}, Snow White~\cite{bentov2016snow}) or BFT protocols (e.g., Algorand~\cite{gilad2017algorand}).
In particular, a combination of PoS with BFT represents a promising approach, which takes advantages of both lottery and voting (i.e., scalability and throughput), while no resources are wasted.

\subsubsection{Pros}
The main feature of PoS protocols, as compared to PoR, is their energy efficiency.
Although some PoS protocols are often combined with a PoR technique (e.g., \cite{bentov2016snow,peercoin}), the overall energy spent is much less than in the case of pure PoR protocols. 

\subsubsection{Cons}
Introduction of PoS protocols has brought PoS specific issues and attacks, while these protocols are still not formally proven to be secure. 
Next, PoS protocols are semi-permissionless -- a node needs to first obtain a stake from any of existing nodes to join the protocol.

\subsubsection{Security Threats and Mitigations}\label{sec:PoS-threats}

\begin{compactitem}    
    \item [\textbf{Nothing-at-Stake:}]
    Since generating a block in PoS does not cost any energy, a node can extend two or more conflicting blocks without risking its stake, and hence increase a chance to be rewarded. 
    Such behavior increases the number of forks and thus time to finality.
    \textit{Countermeasures:}
    Deposit-based solutions (e.g., \cite{buterin2017casper}) require nodes to make a deposit during some fixed period/round and checkpoint-based solutions (e.g., \cite{buterin2017casper,peercoin-wiki,daian2017snow})  employ ``state freezing'' at periodic snapshots of the blockchain, while the blockchain can be reversed maximally up to the recent checkpoint.
    Another option is to use cryptographic solutions~\cite{li2017securing} for revealing identity and a private key of a node that signs two conflicting blocks.
    Another countermeasure is to use backward penalization of nodes that produced two or more conflicting chains~\cite{daian2017snow,buterin2017casper}.
    Finally, PoS protocols can be combined with BFT approaches, and thus forks can hardly occur (e.g.,~\cite{gilad2017algorand}). 
     
    \item [\textbf{Grinding Attack:}]
    If the leader or committee producing a block is determined before the round starts, then the attacker can bias this process to increase his chances of being selected in future.
    For example, if a PoS protocol takes only a hash of the previous block for the election process, the leader of a  block may bias a hash value by suitably adjusting the content of the block in a few attempts.
    \textit{Countermeasures:}
    A grinding attack can be prevented by performing a fresh leader election by an interaction of nodes (e.g., the secure multiparty coin flipping protocol~\cite{kiayias2017ouroboros}) or by private checking whether the output of a verifiable random function (VRF) is below a certain stake-specific threshold (e.g.,~\cite{gilad2017algorand}).
    The input of the VRF is the user's private key and the randomness unambiguously bound to the previous block; hence each consensus node computes the only VRF output during each round.
   
    \item [\textbf{Denial of Service on a Leader/Committee:}]
    If a leader or a committee are publicly determined before the round starts~\cite{kiayias2017ouroboros}, then the adversary may conduct a DoS attack against them and thus cause a restart of the round -- this might be repeated until adversary's desired nodes are elected.
    \textit{Countermeasures:}
    A prevention technique was proposed in Algorand~\cite{gilad2017algorand} -- a node privately determines whether it is a potential leader (or committee member), and immediately releases a block candidate (or a vote) -- hence, after publishing this data, it is too late for a DoS attack.
    The concept of VRF approach was also utilized in other protocols (e.g.,~\cite{david2018ouroboros,hanke2018dfinity}).    
    
    \item [\textbf{Long-Range Attack:}]
    In this attack~\cite{buterin2014long} (a.k.a., posterior corruption~\cite{daian2017snow}), an adversary can ``bribe'' previously influential consensus nodes to sell their private keys or steal the private keys by other means. 
    Since consensus nodes may exchange their crypto-tokens for fiat money anytime, selling their keys impose no expenses and risk.
    If the attacker accumulates keys with enough stake in the past, he may rerun the consensus protocol and rewrite the history of the blockchain. 
    A variant of long-range attack that considers transaction fee-based rewarding and infrequent or no check-points is denoted as a \textit{stake-bleeding} attack~\cite{gavzi2018stake}.
    \textit{Countermeasures:}
    One mitigation is to lock the deposit for a longer time than the period of participation in the consensus~\cite{bano2017consensus}.
    The next mitigation technique is frequent periodic check-pointing, which causes the irreversibility of the blockchain with respect to the last checkpoint.
    Another option is to apply key-evolving cryptography~\cite{franklin2006survey} and forward-secure digital signatures~\cite{bellare1999forward}, which require users to evolve their private keys, while already used keys are erased~\cite{david2018ouroboros}.
    Hence, signatures cannot be forged in the case of compromise.
    The third mitigation technique is enforcing a chain density in a time-domain~\cite{gavzi2018stake} for the protocols where the expected number of participants in each round is known (e.g.,~\cite{kiayias2017ouroboros}).
    The last mitigation technique is context-sensitive transactions, which put the hash of a recent valid block into a transaction itself~\cite{gavzi2018stake}.

\end{compactitem}

%% file: sec/smart_contracts.tex
This layer is responsible for the interpretation of transactions and concerning security threats are related to the privacy of users, confidentiality of data, and smart contract-specific bugs (involving bugs in code and compilers).

\subsection{Transaction Protection}
Mostly, transactions containing plain-text data are digitally signed by private keys of users~\cite{nakamoto2008bitcoin,ethereum-white-paper}, enabling anybody to verify the validity of transactions by corresponding public keys.     
However, such an approach provides only pseudonymous identities that can be traced to real identities, and moreover, it does not ensure confidentiality of data~\cite{feng2019}.

\subsubsection{Security Threats and Countermeasures}\label{sec:privacy-threats}
\begin{compactdesc}
    
    \item[\textbf{Privacy Threats to User Identity.}]
    In most of the blockchains, user identities can be linked with their transactions by various deanonymization techniques, such as network flow analysis, address clustering, transaction fingerprinting~\cite{feng2019,biryukov2014deanonymisation,pustogarov2015deanonymisation}.
     Moreover, blockchains designed with anonymity and privacy features (e.g., Zcash, Monero) are also vulnerable to a few attack strategies~\cite{kappos2018empirical,moser2018empirical}.
    \textit{Countermeasures:}
    Various means are used for obfuscation of user identities, including centralized~\cite{maxwell2013coinjoin,bonneau2014mixcoin} and decentralized~\cite{ruffing2014coinshuffle,bissias2014,ziegeldorf2018} mixing services, ring signatures~\cite{saberhagen2013ring,noether2015ring}, and non-interactive zero-knowledge proofs (NIZKs)~\cite{sasson2014zerocash}.
    Some mixers enable internal linkability by involved parties~\cite{maxwell2013coinjoin} or linkability by the mixers~\cite{bonneau2014mixcoin}, which are also potential threats.
    Unlinkability for all parties can be achieved by multi-party computation~\cite{ziegeldorf2018}, blinding signatures~\cite{valenta2015}, or layered encryption~\cite{ruffing2014coinshuffle}. 
    Ring signatures~\cite{rivestring} provide unlinkability to users in a signing group~\cite{noether2015ring,saberhagen2013ring}, enabling only verification of correctness of a signature, without revealing an identity of a signer.
    
    \item[\textbf{Privacy of data.}]
    NIZKs~\cite{sasson2014zerocash,espel2017proposal} and blind signatures~\cite{heilman2016blindly,valenta2015} can be used for preservation of data privacy. 
    Another method is homomorphic encryption~\cite{pascal99,pedersen91}, which enables to compute some operations over encrypted messages.
    Privacy and confidentiality for smart contract platforms can be achieved through trusted transaction managers~\cite{kosba2016hawk}, trusted hardware~\cite{cheng2018ekiden}, and secure multi-party computations~\cite{zyskind2015enigma}.
\end{compactdesc} 

\subsection{Smart Contracts}
Smart contracts, introduced to automate legal contracts~\cite{szabo1997idea}, now serve as a method for building decentralized applications on blockchains. 
They are usually written in a blockchain-specific programming language that may be Turing-complete and contain arbitrary programming logic or only serve for limited purposes.
In the following, we describe these two contrasting types of smart contract languages. 

\subsubsection{Security Threats and Countermeasures}    
\begin{compactdesc}
    \item[\textbf{Turing-Complete Languages.}] 
    An important aspect of this language category is a large attack surface due to the possibility of arbitrary programming logic.
    Examples of this category are Serpent and Solidity, while Solidity is the most popular and widely-used one.
	\textit{Serpent}~\cite{serpent} is a high-level language that was designed to be simple and similar to the Python language. 
    However, Serpent was designed in untyped fashion, lacking out-of-bound access checks of arrays and accepting invalid code by compilers~\cite{serpentAudit}, which opened the door for plenty of vulnerabilities. 
    Hence, Serpent showed to be as an unsuccessful attempt to simplify the coding phase. 
    \textit{Solidity}~\cite{Solidity} is an object-oriented statically-typed language that is primarily used by Ethereum platform. 
    Contracts written in Solidity can contain various types of vulnerabilities~\cite{atzei2017survey,SoliditySecurityList}, which resulted in many incidents in the past. 
    Mitigations of such vulnerabilities can be done by code analysis tools~\cite{parizi2018empirical,tikhomirov2018smartcheck}, respecting best practices~\cite{SmartContractSecurity,trailofbits}, utilizing known design patterns~\cite{wohrer2018smart}, audits, and testing.
    Various approaches are used for source code analysis, such as linters~\cite{tikhomirov2018smartcheck,solhint,solium}, fuzzers~\cite{jiang2018contractfuzzer}, semantic-based program verifiers~\cite{hildenbrandt2017kevm}, 
    and other symbolic code analyzers~\cite{tsankov2018securify} often using control flow-graph techniques.
    Note that source code of contracts is often not public in contrast to their bytecode.
    For this reason, bytecode decompilers~\cite{zhou2018erays,suiche2017porosity}, analyzers~\cite{nikolic2018finding}, and automated exploit generators~\cite{krupp2018teether} can be utilized. 
    \smallskip

    \item[\textbf{Turing-Incomplete Languages.}]
    The main pro of this category is its design-oriented goal of small attack surface and emphasis on safety but at the cost of limited expressiveness.
    Examples of this category are Pact, Scilla, Vyper.
    \textit{Pact}~\cite{popejoy2016pact} is a declarative language intended for Kadena blockchain and provides type inference and module-guarded tables to prevent direct access to the module. 
    Pact is equipped with the ability to express and check properties of its programs, also leveraging SMT solvers.
    \textit{Scilla}~\cite{sergey2018scilla} is designed to achieve expressiveness and tractability while enabling formal reasoning about contract behavior.
    Every computation utilizes the automata-based model, and computations are realized as standalone atomic transitions that strictly terminate.
    Scilla enables external calls only as the last instruction of a contract, which simplifies proving safety and thus mitigates a few vulnerabilities.
    \textit{Vyper}~\cite{vyper} is an experimental language designed to ease the audit of smart contracts and increase security -- it contains strong typing and bounds/overflows checks.

\end{compactdesc}

%% file: sec/apps.tex
The application layer contains end-user services and applications that are built on top of blockchains; therefore the security threats are specific to particular types of applications.
In the following, we elaborate on common application types.

\subsection{Crypto-Tokens \& Wallets}
Besides cryptocurrencies that provide native crypto-tokens, there are other blockchain applications using crypto-tokens for the purpose of providing owners with rights against the third party (i.e., counterparty tokens) or with a possibility of transferring asset ownership (i.e.,  ownership tokens)~\cite{BCPframework}. 
All types of tokens require the protection of private keys and secrets linked with user identities.
For this purpose, two main categories of wallets have emerged -- \textit{self-sovereign wallets} and \textit{hosted wallets}~\cite{eskandari2018first,2015-Bitcoin-SOK,homoliak2018air}.
Beside technical risks, all crypto-tokens are exposed to regulatory risk, while non-native tokens are in addition exposed to legal risks~\cite{BCPframework}.

\paragraph{Self-Sovereign Wallets}
Users of self-sovereign wallets locally store their private keys and directly interact with the blockchain platform using the keys to sign transactions. 
The instances of these wallets differ in several aspects. 
One of them is isolation of the keys -- there are software wallets that store the keys within the user PC (e.g., Bitcoin Core, Electrum Wallet, MyEtherWallet) as well as hardware wallets that store keys in a sealed storage, while they expose only signing functionality (e.g., Trezor,  Le\-dger, KeepKey, BitLox, CoolBitX).
Another type of wallets enables to customize functionality and security by a smart contract (e.g., TrezorMultisig2of3~\cite{TrezorMultisig2of3}, Ethereum MultiSigWallet~\cite{ConsenSys-wallet}).

\paragraph{Hosted Wallets}
Hosted wallets require a centralized party to provide an interface for interaction with the wallet and thus blockchain.
If a hosted wallet has full control over private keys, it is
referred to as a \textit{server-side wallet} (e.g., Coinbase, Circle Pay Wallet, Luno Wallet), while in the case of keys stored in the users' browsers, the wallets are referred to as \textit{client-side wallets} (e.g., Blockchain Wallet, BTC Wallet, Mycelium Wallet, CarbonWallet, Citowise Wallet).
We refer the reader to works~\cite{homoliak2018air,eskandari2018first} for a security overview of miscellaneous wallet solutions. 

\mysubsubsection{Security Threats and Mitigations}
Since server-side wallets accounted for several compromises~\cite{2018-coindesk-bithumb}, \cite{2014-Mt-Gox}, \cite{2016-Bitfinex-hack}, their popularity have attenuated in favor of client-side wallets.
Client-side wallets do not expose private keys to a centralized party, but they still trust in the online interface provided by such a party, and moreover, their availability is dependent on such a party.
Contrary, self-sovereign wallets do not trust in a third party nor rely on its availability.
However, these wallets are susceptible to key theft (i.e., malware~\cite{2015-CCSM-SecureWorks}, keyloggers~\cite{2015-Bitcoin-SOK,2017-keylogger-bc-malware}).
Possible mitigation of these attacks are hardware wallets displaying details of transactions to the user, while the user confirms signing by a button (e.g., Trezor,  Le\-dger, KeepKey).
Another option is to protect self-sovereign wallets by multi-factor/(-step) authentication using multi-signatures~\cite{TrezorMultisig2of3,ConsenSys-wallet}, threshold-based cryptography~\cite{goldfeder2015securing}, or air-gapped OTPs~\cite{homoliak2018air}.

\subsection{Oracles}
Oracles are trusted entities that provide data reflecting the state of the world beyond the blockchain.
\textit{Prediction markets} (e.g., Augur~\cite{peterson2015augur}, Gnosis~\cite{team2017gnosis}) were created for the purpose of trading the outcome of events -- individuals are incentivized to accurately wager on these outcomes, which serve as data feeds.
\textit{Dedicated data feeds} build on existing blockchain platforms (e.g., PDFS~\cite{guarnizo2018pdfs}, Oraclize~\cite{oraclize}) or create dedicated oracle networks (e.g., ChainLink~\cite{ellisdecentralized}, Witnet~\cite{de2017witnet}) that internally run consensus protocol.

\mysubsubsection{Security Threats}
The data provision time of prediction markets may be long for many applications and the provided set of data events may be also limited.
In contrast, dedicated data feeds enrich a data domain and significantly shorten a provision time; however, they often rely on a trusted party~\cite{guarnizo2018pdfs,oraclize}, which may misbehave or accidentally produce wrong data.
Oracle networks eliminate trust in a single party by a consensus of the group; however, threats related to the consensus layer of this functionality also needs to be considered.
Moreover, for providers that offer authenticated data feeds using trusted hardware~\cite{oraclize,zhang2016town}, a vulnerability in trusted hardware may result in a compromise of the entire data feed.

\subsection{Decentralized Filesystems (DFs)}
DFs serve as a data storage infrastructure running native blockchains (e.g., Storj~\cite{wilkinson2014storj}, Filecoin~\cite{filecoin}, Permacoin~\cite{miller2014permacoin}).
DFs borrow ideas from peer-to-peer file storage systems, but they additionally incentivize data preservation by crypto-tokens. 
Alternatively to native DFs, decoupling of the stored data from the blockchain data is also possible in a few forms of integration with existing blockchains. 
Beside na\"{i}ve storage of integrity proofs to off-chain data,  cloud services (e.g., Amazon Web Services, Google Cloud, IBM), and distributed hash tables (DHT)~\cite{li2018blockchain} are promising approaches.
   
\mysubsubsection{Security Threats and Mitigations}
While native DFs handle availability and decentralization using consensus layer mechanisms, cloud services and DHT solutions rely on a provider's infrastructure and dedicated file sharing networks, respectively.
Sybil attacks claiming redundant storage of the same piece of data can be prevented by unique encryption of each data copy~\cite{wilkinson2014storj}, which, however, puts higher distribution overhead on clients.
Another attack might target the reputation of the network by dropping data and its redundant copies.
A simple mitigation technique is to use multiple consensus nodes for a file upload, which diminishes chances of the attack being successful.
Next mitigation is to hide the number of redundant copies using erasure encoding~\cite{wilkinson2014storj}.

%% file: sec/conclusion.tex
In this paper, we focused on the systematization of knowledge about security aspects of blockchain systems.
We proposed security reference architecture as a stacked model, which we further projected into a threat-risk assessment model that presents various threats and countermeasures.
The proposed stack model consists of four layers: (1) network layer, (2) consensus layer, (3) replicated state machine layer, and (4) application layer.
At each of the layers, we surveyed specific security issues and mitigation techniques.
In future work, we plan to amend the security issues of each layer by details and evidence about real-world incidents.

%% file: sec/ack.tex
This work was supported by the National Research Foundation (NRF), 
Prime Minister’s Office, Singapore, under its National Cybersecurity R\&D Programme (Award No. NRF2016NCR-NCR002-028) and administered by the National Cybersecurity R\&D Directorate.
Also, we would like to thank Pieter Hartel, Daniel Reijsbergen, and Stefanos Leonardos for their valuable feedback.

%% file: ms.bbl
\begin{thebibliography}{100}

\bibitem{akamai2017}
Akamai.
\newblock {Q1 2017 State of the Internet/Connectivity Report}.
\newblock Technical report, 2017.

\bibitem{apostolaki2018sabre}
M.~Apostolaki, G.~Marti, J.~M{\"u}ller, and L.~Vanbever.
\newblock Sabre: Protecting bitcoin against routing attacks.
\newblock 2018.

\bibitem{apostolaki2017hijacking}
M.~Apostolaki, A.~Zohar, and L.~Vanbever.
\newblock Hijacking bitcoin: Routing attacks on cryptocurrencies.
\newblock In {\em IEEE SP}, 2017.

\bibitem{hybriddos}
E.~Arazi.
\newblock Choosing the right ddos solution (part 4): Hybrid protection.
\newblock
  \url{https://blog.radware.com/security/2018/04/choosing-the-right-ddos-solution-hybrid-protection/},
  2018.

\bibitem{insidercorp}
W.~Ashford.
\newblock Corporate networks vulnerable to insider attacks, report finds.
\newblock
  \url{https://www.computerweekly.com/news/252444419/Corporate-networks-vulnerable-to-insider-attacks-report-finds},
  2018.

\bibitem{atzei2017survey}
N.~Atzei, M.~Bartoletti, and T.~Cimoli.
\newblock A survey of attacks on ethereum smart contracts (sok).
\newblock In {\em POST}, 2017.

\bibitem{aublin2013rbft}
P.-L. Aublin, S.~B. Mokhtar, and V.~Qu{\'e}ma.
\newblock {RBFT: Redundant byzantine fault tolerance}.
\newblock In {\em ICDCS}, 2013.

\bibitem{bano2017consensus}
S.~Bano, A.~Sonnino, M.~Al-Bassam, S.~Azouvi, P.~McCorry, S.~Meiklejohn, and
  G.~Danezis.
\newblock Consensus in the age of blockchains.
\newblock {\em arXiv preprint arXiv:1711.03936}, 2017.

\bibitem{bellare1999forward}
M.~Bellare and S.~K. Miner.
\newblock A forward-secure digital signature scheme.
\newblock In {\em CRYPTO'99}, 1999.

\bibitem{bentov2016cryptocurrencies}
I.~Bentov, A.~Gabizon, and A.~Mizrahi.
\newblock Cryptocurrencies without proof of work.
\newblock In {\em FC}, 2016.

\bibitem{bentov2014proof}
I.~Bentov, C.~Lee, A.~Mizrahi, and M.~Rosenfeld.
\newblock Proof of activity: Extending bitcoin's proof of work via proof of
  stake.
\newblock 2014.

\bibitem{bentov2016snow}
I.~Bentov, R.~Pass, and E.~Shi.
\newblock Snow white: Provably secure proofs of stake.
\newblock 2016.

\bibitem{bessani2014state}
A.~Bessani, J.~Sousa, and E.~E. Alchieri.
\newblock State machine replication for the masses with bft-smart.
\newblock In {\em IEEE/IFIP DSN}, 2014.

\bibitem{biryukov2014deanonymisation}
A.~Biryukov, D.~Khovratovich, and I.~Pustogarov.
\newblock Deanonymisation of clients in bitcoin p2p network.
\newblock In {\em ACM CCS}, 2014.

\bibitem{BiryukovP14}
A.~Biryukov and I.~Pustogarov.
\newblock Bitcoin over tor isn't a good idea.
\newblock {\em CoRR}, abs/1410.6079, 2014.

\bibitem{bissias2014}
G.~Bissias, A.~P. Ozisik, B.~N. Levine, and M.~Liberatore.
\newblock Sybil-resistant mixing for bitcoin.
\newblock In {\em WPES}, New York, NY, USA, 2014.

\bibitem{bitcoinj}
{Bitcoinj team}.
\newblock Bitcoinj security model.
\newblock \url{https://bitcoinj.github.io/security-model}, 2019.

\bibitem{weaknessesdos}
BitcoinWiki.
\newblock Weaknesses.
\newblock
  \url{https://en.bitcoin.it/wiki/Weaknesses#Denial_of_Service_.28DoS.29_attacks},
  24 July 2018.

\bibitem{boneh2001short}
D.~Boneh, B.~Lynn, and H.~Shacham.
\newblock Short signatures from the weil pairing.
\newblock In {\em ASIACRYPT}, 2001.

\bibitem{2015-Bitcoin-SOK}
J.~Bonneau, A.~Miller, J.~Clark, A.~Narayanan, J.~A. Kroll, and E.~W. Felten.
\newblock Sok: Research perspectives and challenges for bitcoin and
  cryptocurrencies.
\newblock In {\em IEEE SP}, 2015.

\bibitem{bonneau2014mixcoin}
J.~Bonneau, A.~Narayanan, A.~Miller, J.~Clark, J.~A. Kroll, and E.~W. Felten.
\newblock Mixcoin: Anonymity for bitcoin with accountable mixes.
\newblock In {\em FC}, pages 486--504. Springer, 2014.

\bibitem{timejacking}
A.~Boverman.
\newblock Timejacking \& bitcoin, 2011.

\bibitem{openness2016}
S.~Box and J.~K. West.
\newblock Economic and social benefits of internet openness.
\newblock \url{https://ssrn.com/abstract=2800227}, 22 June 2016.

\bibitem{peercoin-wiki}
Bticoinwiki.
\newblock Peercoin, 2019.

\bibitem{buchman2018tendermint}
E.~Buchman, J.~Kwon, and Z.~Milosevic.
\newblock The latest gossip on bft consensus.
\newblock 2018.

\bibitem{buterin2014long}
V.~Buterin.
\newblock Long-range attacks: The serious problem with adaptive proof of work.
\newblock {\em Ethereum Blog, May}, 2014.

\bibitem{vyper}
V.~Buterin.
\newblock {Vyper}, 2017.

\bibitem{buterin2017casper}
V.~Buterin and V.~Griffith.
\newblock Casper the friendly finality gadget.
\newblock 2017.

\bibitem{cachin2009ByzantinePaxos}
C.~Cachin.
\newblock Yet another visit to paxos.
\newblock {\em IBM Research, Zurich, Switzerland, Tech. Rep. RZ3754}, 2009.

\bibitem{cachin2005random}
C.~Cachin, K.~Kursawe, and V.~Shoup.
\newblock Random oracles in constantinople: Practical asynchronous byzantine
  agreement using cryptography.
\newblock {\em Journal of Cryptology}, 18(3), 2005.

\bibitem{cachin2002sintra}
C.~Cachin and J.~A. Poritz.
\newblock Secure intrusion-tolerant replication on the internet.
\newblock In {\em DSN}, 2002.

\bibitem{cachin2017blockchain}
C.~Cachin and M.~Vukoli{\'c}.
\newblock Blockchain consensus protocols in the wild.
\newblock 2017.

\bibitem{chainalysis2015}
G.~Caffyn.
\newblock Chainalysis ceo denies sybil attack on bitcoin network.
\newblock
  \url{https://www.coindesk.com/chainalysis-ceo-denies-launching-sybil-attack-on-bitcoin-network},
  2015.

\bibitem{castro1999practical}
M.~Castro, B.~Liskov, et~al.
\newblock Practical byzantine fault tolerance.
\newblock In {\em OSDI}, volume~99, 1999.

\bibitem{cheng2018ekiden}
R.~Cheng, F.~Zhang, J.~Kos, W.~He, N.~Hynes, N.~Johnson, A.~Juels, A.~Miller,
  and D.~Song.
\newblock Ekiden: A platform for confidentiality-preserving, trustworthy, and
  performant smart contract execution.
\newblock {\em arXiv preprint arXiv:1804.05141}, 2018.

\bibitem{cc2017}
{Common Criteria}.
\newblock Common criteria for information technologysecurity evaluation. part
  1: Introduction and general model.
\newblock Technical report, 2017.

\bibitem{oraclize}
{Concur Technologies, Inc}.
\newblock Oraclize documentation.
\newblock
  \url{https://github.com/kadena-io/pact/blob/ac759c0882d97b60473cfbb5853b1c25259e1213/docs/pact-properties.md},
  2008.

\bibitem{ConsenSys-wallet}
{Consensys team}.
\newblock {Ethereum MultiSigWallet}, 2017.

\bibitem{conti2018survey}
M.~Conti, E.~S. Kumar, C.~Lal, and S.~Ruj.
\newblock A survey on security and privacy issues of bitcoin.
\newblock {\em IEEE Communications Surveys \& Tutorials}, 20(4), 2018.

\bibitem{crispin2001}
K.~Crispin.
\newblock Alt-roots, alt-tlds.
\newblock IETF Draft, 2001.

\bibitem{daian2017snow}
P.~Daian, R.~Pass, and E.~Shi.
\newblock Snow white: Robustly reconfigurable consensus and applications to
  provably secure proofs of stake.
\newblock In {\em Iacr}. 2017.

\bibitem{david2018ouroboros}
B.~David, P.~Ga{\v{z}}i, A.~Kiayias, and A.~Russell.
\newblock Ouroboros praos: An adaptively-secure, semi-synchronous
  proof-of-stake blockchain.
\newblock In {\em EUROCRYPT}, 2018.

\bibitem{de2017witnet}
A.~S. de~Pedro, D.~Levi, and L.~I. Cuende.
\newblock Witnet: A decentralized oracle network protocol.
\newblock 2017.

\bibitem{2015-CCSM-SecureWorks}
{Dell SecureWorks}.
\newblock Cryptocurrency-stealing malware landscape, 2015.

\bibitem{douceur2002sybil}
J.~R. Douceur.
\newblock The sybil attack.
\newblock In {\em IPTPS}, 2002.

\bibitem{solium}
R.~Dua.
\newblock Solium documentation release 1.0.0.
\newblock \url{https://media.readthedocs.org/pdf/solium/latest/solium.pdf}, 11
  Feb 2019.

\bibitem{duan2014bchain}
S.~Duan, H.~Meling, S.~Peisert, and H.~Zhang.
\newblock Bchain: Byzantine replication with high throughput and embedded
  reconfiguration.
\newblock In {\em OPODIS}, 2014.

\bibitem{Dziembowski_proofsof}
S.~Dziembowski, S.~Faust, V.~Kolmogorov, and K.~Pietrzak.
\newblock Proofs of space.
\newblock In {\em CRYPTO'15}, 2015.

\bibitem{ellisdecentralized}
S.~Ellis, A.~Juels, and S.~Nazarov.
\newblock Chainlink: A decentralized oracle network.

\bibitem{eskandari2018first}
S.~Eskandari, J.~Clark, D.~Barrera, and E.~Stobert.
\newblock A first look at the usability of bitcoin key management.
\newblock 2018.

\bibitem{espel2017proposal}
T.~Espel, L.~Katz, and G.~Robin.
\newblock Proposal for protocol on a quorum blockchain with zero knowledge.
\newblock 2017, 2017.

\bibitem{bind2019}
S.~G. et~al.
\newblock Bind 9 security vulnerability matrix.
\newblock \url{https://kb.isc.org/docs/aa-00913}, 2019.

\bibitem{Solidity}
{Ethereum team}.
\newblock Solidity.
\newblock \url{https://solidity.readthedocs.io/en/v0.5.4/index.html#}.

\bibitem{ethereum-white-paper}
{Ethereum team}.
\newblock {A Next-Generation Smart Contract and Decentralized Application
  Platform}.
\newblock
  \url{https://github.com/ethereum/wiki/wiki/White-Paper\#modified-ghost-implementation},
  2018.

\bibitem{serpent}
{Etherum team}.
\newblock {Serpent}, 2017.

\bibitem{eyal2018majority}
I.~Eyal and E.~G. Sirer.
\newblock Majority is not enough: Bitcoin mining is vulnerable.
\newblock {\em Communications of the ACM}, 61(7), 2018.

\bibitem{feng2019}
Q.~Feng, D.~He, S.~Zeadally, M.~K. Khan, and N.~Kumar.
\newblock A survey on privacy protection in blockchain system.
\newblock {\em Journal of Network and Computer Applications}, 126, 2019.

\bibitem{franklin2006survey}
M.~Franklin.
\newblock A survey of key evolving cryptosystems.
\newblock {\em International Journal of Security and Networks}, 1(1-2), 2006.

\bibitem{gavzi2018stake}
P.~Ga{\v{z}}i, A.~Kiayias, and A.~Russell.
\newblock Stake-bleeding attacks on proof-of-stake blockchains.
\newblock In {\em IEEE CVCBT}, 2018.

\bibitem{gilad2017algorand}
Y.~Gilad, R.~Hemo, S.~Micali, G.~Vlachos, and N.~Zeldovich.
\newblock Algorand: Scaling byzantine agreements for cryptocurrencies.
\newblock In {\em SOSP}, 2017.

\bibitem{team2017gnosis}
{Gnosis Team}.
\newblock Gnosis-whitepaper.
\newblock {\em URL: https://gnosis.
  pm/resources/default/pdf/gnosis\_whitepaper. pdf}, 2017.

\bibitem{goldfeder2015securing}
S.~Goldfeder, R.~Gennaro, H.~Kalodner, J.~Bonneau, J.~A. Kroll, E.~W. Felten,
  and A.~Narayanan.
\newblock Securing bitcoin wallets via a new dsa/ecdsa threshold signature
  scheme, 2015.

\bibitem{guarnizo2018pdfs}
J.~Guarnizo and P.~Szalachowski.
\newblock Pdfs: Practical data feed service for smart contracts.
\newblock 2018.

\bibitem{hanke2018dfinity}
T.~Hanke, M.~Movahedi, and D.~Williams.
\newblock Dfinity technology overview series, consensus system.
\newblock 2018.

\bibitem{heilman2016blindly}
E.~Heilman, F.~Baldimtsi, and S.~Goldberg.
\newblock Blindly signed contracts: Anonymous on-blockchain and off-blockchain
  bitcoin transactions.
\newblock In {\em FC}, 2016.

\bibitem{heilman2015eclipse}
E.~Heilman, A.~Kendler, A.~Zohar, and S.~Goldberg.
\newblock Eclipse attacks on bitcoin's peer-to-peer network.
\newblock In {\em USENIX Security}, 2015.

\bibitem{minerddos}
S.~Higgins.
\newblock Bitcoin mining pools targeted in wave of ddos attacks.
\newblock \url{https://www.coindesk.com/bitcoin-mining-pools-ddos-attacks},
  2015.

\bibitem{hildenbrandt2017kevm}
E.~Hildenbrandt, M.~Saxena, X.~Zhu, N.~Rodrigues, P.~Daian, D.~Guth, and
  G.~Rosu.
\newblock Kevm: A complete semantics of the ethereum virtual machine.
\newblock Technical report, 2017.

\bibitem{homoliak2018air}
I.~Homoliak, D.~Breitenbacher, A.~Binder, and P.~Szalachowski.
\newblock An air-gapped 2-factor authentication for smart-contract wallets.
\newblock 2018.

\bibitem{homoliak2019insight}
I.~Homoliak, F.~Toffalini, J.~Guarnizo, Y.~Elovici, and M.~Ochoa.
\newblock Insight into insiders and it: A survey of insider threat taxonomies,
  analysis, modeling, and countermeasures.
\newblock {\em ACM Computing Surveys (CSUR)}, 52(2):30, 2019.

\bibitem{honsi2017spacemint}
T.~H{\o}nsi.
\newblock Spacemint-a cryptocurrency based on proofs of space.
\newblock Master's thesis, NTNU, 2017.

\bibitem{hyperledger1}
{Hyperledger team}.
\newblock Hyperledger architecture, volume 1: Consensus, 2017.

\bibitem{jiang2018contractfuzzer}
B.~Jiang, Y.~Liu, and W.~Chan.
\newblock Contractfuzzer: Fuzzing smart contracts for vulnerability detection.
\newblock In {\em ASE}, 2018.

\bibitem{kappos2018empirical}
G.~Kappos, H.~Yousaf, M.~Maller, and S.~Meiklejohn.
\newblock An empirical analysis of anonymity in zcash.
\newblock In {\em USENIX Security}, 2018.

\bibitem{kiayias2017ouroboros}
A.~Kiayias, A.~Russell, B.~David, and R.~Oliynykov.
\newblock Ouroboros: A provably secure proof-of-stake blockchain protocol.
\newblock In {\em CRYPTO'17}, 2017.

\bibitem{kogias2016byzcoin}
E.~K. Kogias, P.~Jovanovic, N.~Gailly, I.~Khoffi, L.~Gasser, and B.~Ford.
\newblock Enhancing bitcoin security and performance with strong consistency
  via collective signing.
\newblock In {\em USENIX Security}, 2016.

\bibitem{Kokoris-KogiasJ18-omniledger}
E.~Kokoris{-}Kogias, P.~Jovanovic, L.~Gasser, N.~Gailly, E.~Syta, and B.~Ford.
\newblock Omniledger: {A} secure, scale-out, decentralized ledger via sharding.
\newblock In {\em 2018 {IEEE} {S\&P} 2018, Proceedings, 21-23 May 2018, San
  Francisco, California, {USA}}, 2018.

\bibitem{kosba2016hawk}
A.~Kosba, A.~Miller, E.~Shi, Z.~Wen, and C.~Papamanthou.
\newblock Hawk: The blockchain model of cryptography and privacy-preserving
  smart contracts.
\newblock In {\em IEEE S\&P}, 2016.

\bibitem{krupp2018teether}
J.~Krupp and C.~Rossow.
\newblock teether: Gnawing at ethereum to automatically exploit smart
  contracts.
\newblock In {\em USENIX Security}, 2018.

\bibitem{rfc8205}
M.~Lepinski and K.~Sriram.
\newblock Bgpsec protocol specification.
\newblock RFC 8205, 2017.

\bibitem{hdddos}
S.~D. Lerner.
\newblock New dos vuln by forcing continuous hard disk seek/read activity
  (fixed in 0.8.0).
\newblock \url{https://bitcointalk.org/index.php?topic=144122.0}, 2019.

\bibitem{li2018blockchain}
R.~Li, T.~Song, B.~Mei, H.~Li, X.~Cheng, and L.~Sun.
\newblock Blockchain for large-scale internet of things data storage and
  protection.
\newblock {\em IEEE Transactions on Services Computing}, 2018.

\bibitem{li2017securing}
W.~Li, S.~Andreina, J.-M. Bohli, and G.~Karame.
\newblock Securing proof-of-stake blockchain protocols.
\newblock In {\em DPM}. 2017.

\bibitem{SoliditySecurityList}
A.~Manning.
\newblock Solidity security: Comprehensive list of known attack vectors and
  common anti-patterns.
\newblock \url{https://blog.sigmaprime.io/solidity-security.html}, 2018.

\bibitem{marcus2018low}
Y.~Marcus, E.~Heilman, and S.~Goldberg.
\newblock Low-resource eclipse attacks on ethereum's peer-to-peer network.
\newblock 2018, 2018.

\bibitem{maxwell2013coinjoin}
G.~Maxwell.
\newblock Coinjoin: Bitcoin privacy for the real world.
\newblock In {\em Post on Bitcoin forum}, 2013.

\bibitem{mcknight1995}
L.~W. McKnight and J.~P. Bailey.
\newblock An introduction to internet economics.
\newblock 1995.

\bibitem{feather-forks}
A.~Miller.
\newblock {Feather-forks: enforcing a blacklist with sub - 50\% hash power}.
\newblock \url{https://bitcointalk.org/index.php?topic=312668.0}, 2013.

\bibitem{miller2014permacoin}
A.~Miller, A.~Juels, E.~Shi, B.~Parno, and J.~Katz.
\newblock Permacoin: Repurposing bitcoin work for data preservation.
\newblock In {\em IEEE SP}, 2014.

\bibitem{miller2015nonoutsourceable}
A.~Miller, A.~Kosba, J.~Katz, and E.~Shi.
\newblock Nonoutsourceable scratch-off puzzles to discourage bitcoin mining
  coalitions.
\newblock In {\em ACM CCS}, 2015.

\bibitem{millertopology2015}
A.~Miller, J.~Litton, A.~Pachulski, N.~Gupta, D.~Levin, N.~Spring, and
  B.~Bhattacharjee.
\newblock Discovering bitcoin' s public topology and influential nodes.
\newblock 2015.

\bibitem{miller2016honey}
A.~Miller, Y.~Xia, K.~Croman, E.~Shi, and D.~Song.
\newblock The honey badger of bft protocols.
\newblock In {\em ACM CCS}, 2016.

\bibitem{BCPframework}
MME.
\newblock Conceptual framework for legal and risk assessment of crypto tokens.
\newblock 2018.

\bibitem{moser2018empirical}
M.~M{\"o}ser, K.~Soska, E.~Heilman, K.~Lee, H.~Heffan, S.~Srivastava, K.~Hogan,
  J.~Hennessey, A.~Miller, A.~Narayanan, et~al.
\newblock An empirical analysis of traceability in the monero blockchain.
\newblock {\em PETS}, 2018(3), 2018.

\bibitem{nakamoto2008bitcoin}
S.~Nakamoto.
\newblock Bitcoin: A peer-to-peer electronic cash system, 2008.

\bibitem{nayak2016stubborn}
K.~Nayak, S.~Kumar, A.~Miller, and E.~Shi.
\newblock Stubborn mining: Generalizing selfish mining and combining with an
  eclipse attack.
\newblock In {\em IEEE EuroSP}, 2016.

\bibitem{nikolic2018finding}
I.~Nikoli{\'c}, A.~Kolluri, I.~Sergey, P.~Saxena, and A.~Hobor.
\newblock Finding the greedy, prodigal, and suicidal contracts at scale.
\newblock In {\em ACM ACSAC}, 2018.

\bibitem{noether2015ring}
S.~Noether.
\newblock Ring signature confidential transactions for monero, 2015.

\bibitem{slimcoin}
P4Titan.
\newblock Slimcoin: A peer-to-peer crypto-currency with proof-of-burn.
\newblock Technical report, 2014.

\bibitem{pascal99}
P.~Paillier.
\newblock Public-key cryptosystems based on composite degree residuosity
  classes.
\newblock In J.~Stern, editor, {\em EUROCRYPT'99}, Berlin, Heidelberg, 1999.

\bibitem{parizi2018empirical}
R.~M. Parizi, A.~Dehghantanha, K.-K.~R. Choo, and A.~Singh.
\newblock Empirical vulnerability analysis of automated smart contracts
  security testing on blockchains.
\newblock In {\em CASCON}, 2018.

\bibitem{spacecoin}
S.~Park, K.~Pietrzak, J.~Alwen, G.~Fuchsbauer, and P.~Gazi.
\newblock Spacecoin: A cryptocurrency based on proofs of space.
\newblock Technical report, 2015.

\bibitem{pass2017fruitchains}
R.~Pass and E.~Shi.
\newblock Fruitchains: A fair blockchain.
\newblock In {\em PODC}, 2017.

\bibitem{pedersen91}
T.~P. Pedersen.
\newblock Non-interactive and information-theoretic secure verifiable secret
  sharing.
\newblock In J.~Feigenbaum, editor, {\em CRYPTO'91}, Berlin, Heidelberg, 1992.

\bibitem{peterson2015augur}
J.~Peterson and J.~Krug.
\newblock Augur: a decentralized, open-source platform for prediction markets.
\newblock 2015.

\bibitem{2017-keylogger-bc-malware}
A.~Peyton.
\newblock Cyren sounds siren over bitcoin siphon scam, 2017.

\bibitem{popejoy2016pact}
S.~Popejoy.
\newblock {The Pact smart contract language}, 2016.

\bibitem{filecoin}
{Protocol Labs}.
\newblock Filecoin: A decentralized storage network.
\newblock Technical report, 2017.

\bibitem{solhint}
Protofire.
\newblock Solhint project.
\newblock \url{https://github.com/protofire/solhint}, 2017.

\bibitem{pustogarov2015deanonymisation}
I.~Pustogarov.
\newblock {\em Deanonymisation techniques for Tor and Bitcoin}.
\newblock PhD thesis, University of Luxembourg, 2015.

\bibitem{2014-Mt-Gox}
{Rachel Abrams and Nathaniel Popper}.
\newblock {Trading Site Failure Stirs Ire and Hope for Bitcoin}, 2014.

\bibitem{2016-Bitfinex-hack}
{Reuters}.
\newblock {Bitcoin Worth \$72M Was Stolen in Bitfinex Exchange Hack in Hong
  Kong}, 2016.

\bibitem{rivestring}
R.~L. Rivest, A.~Shamir, and Y.~Tauman.
\newblock How to leak a secret.
\newblock In C.~Boyd, editor, {\em ASIACRYPT}, Berlin, Heidelberg, 2001.

\bibitem{rizun2016subchains}
P.~R. Rizun.
\newblock Subchains: A technique to scale {B}itcoin and improve the user
  experience.
\newblock {\em Ledger}, 1, 2016.

\bibitem{rodrigues2010}
R.~Rodrigues and P.~Druschel.
\newblock Peer-to-peer systems.
\newblock {\em Commun. ACM}, 53(10), 2010.

\bibitem{rosenfeld2011analysis}
M.~Rosenfeld.
\newblock Analysis of {B}itcoin pooled mining reward systems.
\newblock 2011.

\bibitem{internetadversary}
D.~S.~H. Rosenthal, P.~Maniatis, M.~Roussopoulos, T.~J. Giuli, and M.~Baker.
\newblock Notes on the design of an internet adversary.
\newblock {\em CoRR}, cs.DL/0411078, 2004.

\bibitem{ruffing2014coinshuffle}
T.~Ruffing, P.~Moreno-Sanchez, and A.~Kate.
\newblock Coinshuffle: Practical decentralized coin mixing for bitcoin.
\newblock In {\em ESORICS}, 2014.

\bibitem{peercoin}
{S. King and S. Nadal}.
\newblock Ppcoin: Peer-to-peer crypto-currency with proof-of-stake.
\newblock Technical report, 2012.

\bibitem{sapirshtein2016optimal}
A.~Sapirshtein, Y.~Sompolinsky, and A.~Zohar.
\newblock Optimal selfish mining strategies in {B}itcoin.
\newblock In {\em FC}, 2016.

\bibitem{sasson2014zerocash}
E.~B. Sasson, A.~Chiesa, C.~Garman, M.~Green, I.~Miers, E.~Tromer, and
  M.~Virza.
\newblock Zerocash: Decentralized anonymous payments from bitcoin.
\newblock In {\em IEEE SP}, 2014.

\bibitem{schneider1990implementing}
F.~B. Schneider.
\newblock Implementing fault-tolerant services using the state machine
  approach: A tutorial.
\newblock {\em ACM CSUR}, 22(4), 1990.

\bibitem{sergey2018scilla}
I.~Sergey, A.~Kumar, and A.~Hobor.
\newblock Scilla: a smart contract intermediate-level language.
\newblock 2018.

\bibitem{fullnodeddos}
D.~Shares.
\newblock Major {{DDoS}} attacks hit bitcoin.com.
\newblock
  \url{https://news.bitcoin.com/ddos-attacks-bitcoin-com-uncensored-information/},
  2017.

\bibitem{shoup2000practical}
V.~Shoup.
\newblock Practical threshold signatures.
\newblock In {\em EUROCRYPT}, 2000.

\bibitem{SmartContractSecurity}
SmartContractSecurity.
\newblock Smart contract weakness classification registry.
\newblock \url{https://github.com/SmartContractSecurity/SWC-registry/}, 2019.

\bibitem{sompolinsky2013accelerating}
Y.~Sompolinsky and A.~Zohar.
\newblock Accelerating bitcoin's transaction processing. fast money grows on
  trees, not chains.
\newblock 2013(881), 2013.

\bibitem{Son2010DNS}
S.~Son and V.~Shmatikov.
\newblock The hitchhiker's guide to dns cache poisoning.
\newblock In S.~Jajodia and J.~Zhou, editors, {\em SecureComm}, Berlin,
  Heidelberg, 2010.

\bibitem{suiche2017porosity}
M.~Suiche.
\newblock Porosity: A decompiler for blockchain-based smart contracts bytecode.
\newblock {\em DEF CON}, 25, 2017.

\bibitem{swift06}
D.~Swift.
\newblock {A Practical Application ofSIM/SEM/SIEM Automating Threat
  Identification}.
\newblock
  \url{https://www.sans.org/reading-room/whitepapers/logging/practical-application-sim-sem-siem-automating-threat-identification-1781},
  2006).

\bibitem{szabo1997idea}
N.~Szabo.
\newblock The idea of smart contracts.

\bibitem{szalachowski2018short}
P.~Szalachowski.
\newblock (short paper) towards more reliable bitcoin timestamps.
\newblock In {\em IEEE CVCBT}, 2018.

\bibitem{tikhomirov2018smartcheck}
S.~Tikhomirov, E.~Voskresenskaya, I.~Ivanitskiy, R.~Takhaviev, E.~Marchenko,
  and Y.~Alexandrov.
\newblock Smartcheck: Static analysis of ethereum smart contracts.
\newblock In {\em WETSEB}, 2018.

\bibitem{trailofbits}
trailofbits.
\newblock Awesome ethereum security.
\newblock \url{https://github.com/trailofbits/awesome-ethereum-security}, 11
  Aug 2018.

\bibitem{tsankov2018securify}
P.~Tsankov, A.~Dan, D.~Drachsler-Cohen, A.~Gervais, F.~Buenzli, and M.~Vechev.
\newblock Securify: Practical security analysis of smart contracts.
\newblock In {\em ACM CCS}, 2018.

\bibitem{TrezorMultisig2of3}
{Unchained Capital}.
\newblock {TrezorMultisig2of3: Ethereum Multisignature smart contract}, 2018.

\bibitem{valenta2015}
L.~Valenta and B.~Rowan.
\newblock Blindcoin: Blinded, accountable mixes for bitcoin.
\newblock In M.~Brenner, N.~Christin, B.~Johnson, and K.~Rohloff, editors, {\em
  Financial Crypto}, Berlin, Heidelberg, 2015.

\bibitem{saberhagen2013ring}
N.~van Saberhagen.
\newblock Cryptonote v 2.0.
\newblock \url{https://cryptonote.org/whitepaper.pdf}, 2013.

\bibitem{wang2018survey}
W.~Wang, D.~T. Hoang, Z.~Xiong, D.~Niyato, P.~Wang, P.~Hu, and Y.~Wen.
\newblock A survey on consensus mechanisms and mining management in blockchain
  networks.
\newblock 2018.

\bibitem{wilkinson2014storj}
S.~Wilkinson, T.~Boshevski, J.~Brandoff, and V.~Buterin.
\newblock Storj a peer-to-peer cloud storage network.
\newblock 2014.

\bibitem{wohrer2018smart}
M.~Wohrer and U.~Zdun.
\newblock Smart contracts: Security patterns in the ethereum ecosystem and
  solidity.
\newblock In {\em IWBOSE}, 2018.

\bibitem{2018-coindesk-bithumb}
{Wolfie Zhao}.
\newblock {Bithumb \$31 Million Crypto Exchange Hack: What We Know (And
  Don't)}, 2018.

\bibitem{wust2016ethereum}
K.~W{\"u}st and A.~Gervais.
\newblock Ethereum eclipse attacks.
\newblock Technical report, 2016.

\bibitem{ZamaniM018-rapidchain}
M.~Zamani, M.~Movahedi, and M.~Raykova.
\newblock Rapidchain: Scaling blockchain via full sharding.
\newblock In {\em ACM CCS}, 2018.

\bibitem{zamyatinflux}
A.~Zamyatin, N.~Stifter, P.~Schindler, E.~Weippl, and W.~J. Knottenbelt.
\newblock {Flux: Revisiting Near Blocks for Proof-of-Work Blockchains}, 2018.
\newblock \url{https://eprint.iacr.org/2018/415/20180529:172206}.

\bibitem{serpentAudit}
{Zeppelin Sollutions}.
\newblock Serpent compiler audit.
\newblock
  \url{https://docs.google.com/document/d/1_PqXuAkvgUAOG3jbBvaUvqN6W90eJ3N4IdTLNMRAijo/edit#heading=h.pe41jxc4c6xs},
  2017.

\bibitem{zhang2016town}
F.~Zhang, E.~Cecchetti, K.~Croman, A.~Juels, and E.~Shi.
\newblock Town crier: An authenticated data feed for smart contracts.
\newblock In {\em ACM CCS}, 2016.

\bibitem{zhang2017publish}
R.~Zhang and B.~Preneel.
\newblock Publish or perish: A backward-compatible defense against selfish
  mining in bitcoin.
\newblock In {\em CT-RSA}, 2017.

\bibitem{2019-pow-evaluation}
R.~Zhang and B.~Preneel.
\newblock Lay down the common metrics: Evaluating proof-of-work consensus
  protocols' security.
\newblock In {\em IEEE SP}, 2019.

\bibitem{zhou2018erays}
Y.~Zhou, D.~Kumar, S.~Bakshi, J.~Mason, A.~Miller, and M.~Bailey.
\newblock Erays: reverse engineering ethereum's opaque smart contracts.
\newblock In {\em USENIX Security}, 2018.

\bibitem{ziegeldorf2018}
J.~H. Ziegeldorf, R.~Matzutt, M.~Henze, F.~Grossmann, and K.~Wehrle.
\newblock Secure and anonymous decentralized bitcoin mixing.
\newblock {\em Future Generation Computer Systems}, 80, 2018.

\bibitem{zilliqa2017zilliqa}
{ZILLIQA Team}.
\newblock {The ZILLIQA Technical Whitepaper}, 2017.

\bibitem{zyskind2015enigma}
G.~Zyskind, O.~Nathan, and A.~Pentland.
\newblock Enigma: Decentralized computation platform with guaranteed privacy.
\newblock 2015.

\end{thebibliography}
